\documentclass[twocolumn]{aastex631}

\usepackage{textgreek, gensymb, needspace, bm}
\usepackage[]{fixltx2e}
\usepackage[bottom]{footmisc}
\received{March 23, 2023}
\revised{April 4, 2023}
\accepted{April 4, 2023}

\submitjournal{ApJ}

\shorttitle{Properties of G 68-34 A \& B}
\shortauthors{Pass \& Charbonneau}
\begin{document}

\title{G 68-34: A Double-Lined M-Dwarf Eclipsing Binary in a Hierarchical Triple System}

\author[0000-0002-1533-9029]{Emily K. Pass}
\affiliation{Center for Astrophysics $\vert$ Harvard \& Smithsonian, 60 Garden Street, Cambridge, MA 02138, USA}

\author[0000-0002-9003-484X]{David Charbonneau}
\affiliation{Center for Astrophysics $\vert$ Harvard \& Smithsonian, 60 Garden Street, Cambridge, MA 02138, USA}



\begin{abstract}
Using high-resolution spectra from the Tillinghast Reflector Echelle Spectrograph (TRES) and photometry from sector 56 of the Transiting Exoplanet Survey Satellite (TESS), we report that the nearby M dwarf G 68-34 is a double-lined eclipsing binary. The pair is spin--orbit synchronized with a period of 0.655 days. The light curve shows significant spot modulation with a larger photometric amplitude than that of the grazing eclipses. We perform a joint fit to the spectroscopic and photometric data, obtaining masses of $0.3280\pm 0.0034$M$_\odot$ and $0.3207\pm 0.0036$M$_\odot$ and radii of $0.345\pm 0.014$R$_\odot$ and $0.342\pm 0.014$R$_\odot$ after marginalizing over unknowns in the starspot distribution. This system adds to the small but growing population of fully convective M dwarfs with precisely measured masses and radii that can be used to test models of stellar structure. The pair also has a white dwarf primary at 9" separation, with the system known to be older than 5 Gyr from the white-dwarf cooling age. The binarity of G 68-34 confirms our hypothesis from \citet{Pass2022}: in that work, we noted that \hbox{G 68-34} was both rapidly rotating and old, highly unusual given our understanding of the spindown of M dwarfs, and that a close binary companion may be responsible. 
\end{abstract}


\section{Introduction} \label{sec:intro}

In this paper, we discuss a system that was previously identified as a widely separated binary \citep{Luyten1995}. \hbox{LP 463-28} and \hbox{G 68-34} are a pair at 39 pc, consisting of a white dwarf and an M dwarf separated by 9". These distances and separations are based on Gaia astrometry \citep{Gaia2016, Gaia2022}. In \citet{Pass2022}, we found that G 68-34 rotates with a period of 0.655 days, yet the system is older than 5 Gyr based on the white-dwarf cooling age of the primary. Given its anomalously rapid rotation, we hypothesized that \hbox{G 68-34} may be a close binary, which motivated this work. Here, we study G 68-34 with high-resolution spectroscopy, finding that it is indeed a double-lined spectroscopic binary (SB2). Moreover, photometry from the Transiting Exoplanet Survey Satellite (TESS) reveals that the pair also eclipses, making it a double-lined eclipsing binary (DLEB).

It is common for a close stellar binary to be orbited by a widely separated third component \citep{Tokovinin2006}. Such an architecture is thought to be the result of the Kozai-Lidov mechanism, with the presence of the outer companion driving the inner binary towards small separations \citep[e.g.,][]{Fabrycky2007}.

DLEBs such as \hbox{G 68-34 AB} are important systems for testing models of stellar evolution, as they allow masses and radii of the two stars to be measured. To date, the number of fully convective M dwarfs discovered and studied in DLEBs is modest: \citet{Schweitzer2019} identified 22 such stars in the literature with masses less than 0.35M$_\odot$, with only 13 having masses and radii estimated to better than 5\% precision. Study of these systems is of current interest, as their observed radii have been found to be inflated by as much as 10--15\% relative to theoretical models \citep[e.g.,][]{Kesseli2018}. The \hbox{G 68-34} AB system is also particularly noteworthy because we have an age estimate from the white-dwarf primary; in general, it is very difficult to measure the ages of M dwarfs.

In Section~\ref{sec:data}, we present the spectroscopic and photometric data used in our analysis. In Section~\ref{sec:analysis}, we perform a joint fit of these data sets and extract stellar parameters. We discuss our results in Section~\ref{sec:discussion} and conclude in Section~\ref{sec:conclusion}.

\begin{figure*}[t]
    \centering
    \makebox[\textwidth][c]{\includegraphics[width=0.9\textwidth]{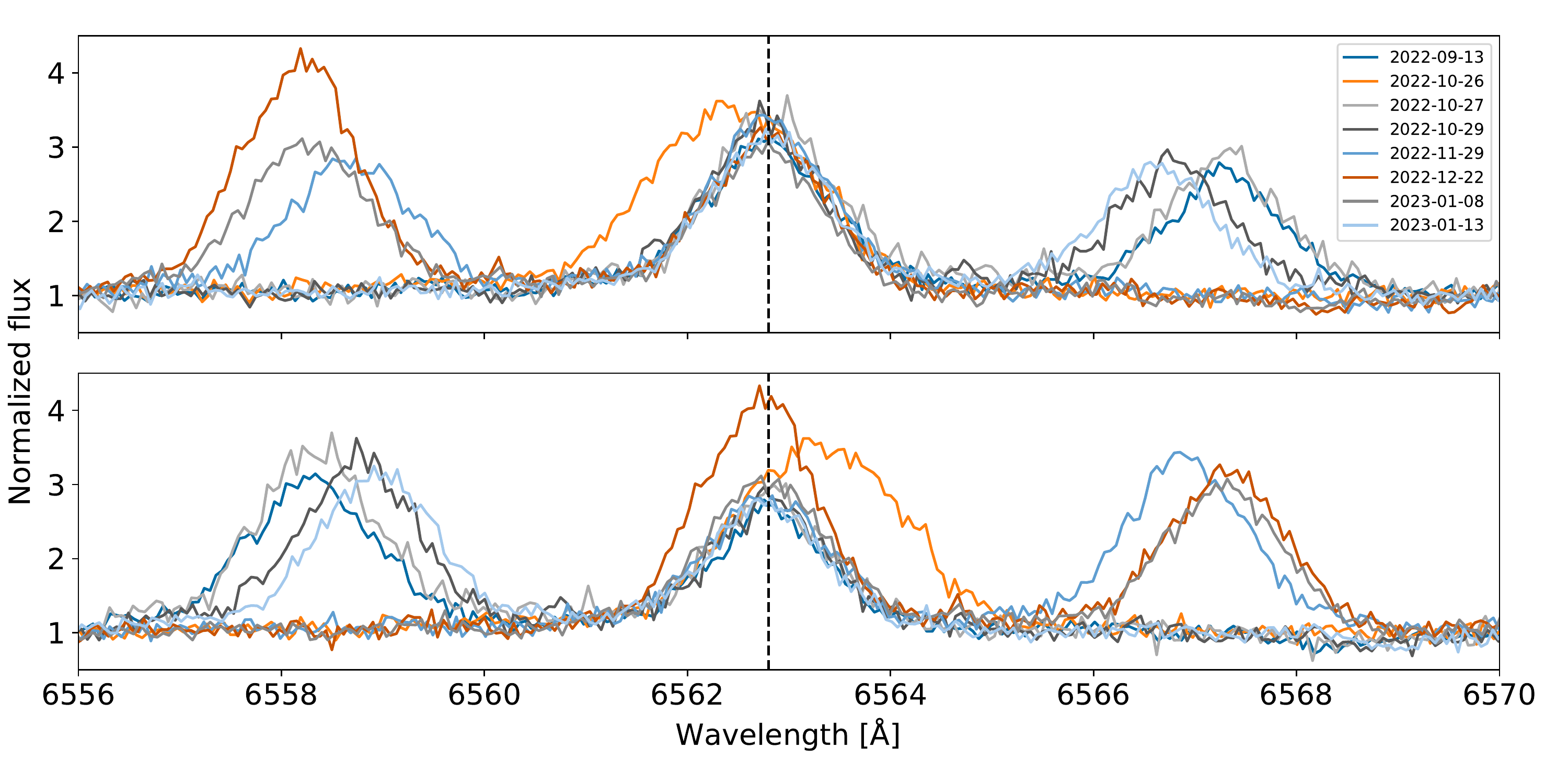}}
    \caption{Both G 68-34 A and B are active in H\textalpha, with two emission lines apparent in each TRES spectrum. In the upper panel, we have shifted the spectra to the rest frame of their brightest set of lines, such that the H\textalpha\ emission for the brighter A component always appears at 6562.8\AA\ (the black dashed line). In the lower panel, we have shifted the spectra to the rest frame of the B component.}
    \label{fig:Ha}
\end{figure*}

\section{Data Collection and Reduction}
\label{sec:data}
\subsection{Spectroscopy}
We observed G 68-34 with the TRES spectrograph ($R$=$44000$; \citealt{Szentgyorgyi2007}) on the \hbox{1.5m} telescope at the Fred Lawrence Whipple Observatory (FLWO). We observed the star at eight epochs between 2022 September and 2023 January, with exposure times varying between 2160 and 3600 seconds depending on sky conditions. Each of the observations consist of three subexposures that we extract separately using the standard TRES pipeline \citep{Buchhave2010}. Two sets of lines are clearly visible in the spectra, as we show in Figure~\ref{fig:Ha} using the H\textalpha\ emission feature at 6563\AA.

We follow the method of \citet{Winters2020} to identify and characterize SB2s from TRES spectra, which is based on the \texttt{TODCOR} technique \citep{Zucker1994}. The analysis uses TRES order 41, corresponding to wavelengths 7065–7165 \AA, which is dominated by TiO bandhead features for M dwarfs. As in \hbox{\citet{Winters2020}}, we use a spectrum of Barnard's Star as our cross-correlation template. The results of this \texttt{TODCOR} analysis are given in Table~\ref{tab:rv}. We measure a light ratio of \hbox{$F_{\rm B}/F_{\rm A}=0.94$} and obtain $v$sin$i$ estimates of \hbox{$v$sin$i_{\rm A} = 24$kms$^{-1}$} and \hbox{$v$sin$i_{\rm B} = 23$kms$^{-1}$}.

Both components show substantial H\textalpha\ emission. In the blended spectra, we measure a median H\textalpha\ equivalent width of -3.66\AA\ for the brighter set of lines and -3.05\AA\ for the dimmer set of lines, neglecting the 2022 October 26 epoch where the H\textalpha\ features of the two components are blended. To measure these equivalent widths, we use the method from \citet{Medina2020} but adjust the continuum regions outwards to 6551.7--6555.6\AA\ and 6569.0--6572.9\AA, such that we avoid including the other emission line in our measurement of the continuum. The equivalent width in the blended spectrum is related to the true width by \hbox{EW$_{\rm true} = (1/X + 1) \times \rm{EW}_{\rm{obs}}$,} where $X$ is the light ratio $F_{\rm target}/F_{\rm companion}$. From the \texttt{TODCOR} analysis, we found a light ratio of $F_{\rm dimmer}/F_{\rm brighter}$=0.94, or $F_{\rm brighter}/F_{\rm dimmer}$=1.06. We therefore correct our equivalent widths to -7.1\AA\ for A and -6.3\AA\ for B. We also note that the 2022 December 22 epoch appears to capture a flare in the B component: we measure heightened H\textalpha\ emission from B, while the emission from A is consistent with the other observations. Specifically, we measure an equivalent width of -4.8\AA\ for B at this epoch, corresponding to a deblended width of -9.8\AA.

\begin{deluxetable}{lrrrr}[t]
\tabletypesize{\footnotesize}
\tablecolumns{7}
\tablewidth{0pt}
 \tablecaption{\texttt{TODCOR} results for G 68-34 A and B \label{tab:rv}}
 \tablehead{
 \colhead{\textbf{BJD}} & 
 \colhead{\textbf{RV\bm{$_{\rm A}$}}} &
 \colhead{\textbf{RV\bm{$_{\rm B}$}}} &
 \colhead{\vspace{-0.1cm} \bm{$h$}} &  
 \colhead{\bm{$t_{\rm exp}$}} \\[-0.2cm]
 \colhead{[d]} &
 \colhead{[kms$^{-1}$]} &
 \colhead{[kms$^{-1}]$} &
 \colhead{} &
 \colhead{[s]}}
\startdata
2459836.7154 & -115.593 & 89.731 & 0.556 & 1200 \\
2459836.7287 & -114.230 & 90.078 & 0.549 & 1000 \\
2459836.7406 & -114.401 & 88.392 & 0.439 & 1000 \\
2459879.7423 & 18.865 & -48.866 & 0.538 & 1200 \\
2459879.7564 & 6.036 & -36.035 & 0.549 & 1200 \\
2459879.7705 & -5.670 & -20.787 & 0.558 & 1200 \\
2459880.5950 & -112.943 & 90.002 & 0.364 & 720 \\
2459880.6043 & -115.352 & 92.374 & 0.383 & 720 \\
2459880.6132 & -111.853 & 89.704 & 0.419 & 720 \\
2459882.5956 & -111.454 & 85.026 & 0.367 & 1000 \\
2459882.6076 & -103.928 & 82.920 & 0.370 & 1000 \\
2459882.6306 & -93.934 & 69.176 & 0.420 & 1000 \\
2459913.7033 & 84.663 & -112.163 & 0.444 & 1000 \\
2459913.7156 & 80.475 & -105.97 & 0.419 & 1000 \\
2459913.7279 & 73.458 & -103.691 & 0.419 & 1000 \\
2459936.5880 & 88.881 & -117.998 & 0.469 & 900 \\
2459936.5990 & 91.046 & -115.626 & 0.500 & 900 \\
2459936.6097 & 87.593 & -117.620 & 0.430 & 900 \\
2459953.6018 & 86.443 & -114.591 & 0.493 & 1200 \\
2459953.6164 & 90.622 & -118.033 & 0.532 & 1200 \\
2459953.6306 & 90.338 & -118.165 & 0.537 & 1200 \\
2459958.5801 & -103.494 & 86.883 & 0.475 & 1200 \\
2459958.5942 & -98.848 & 75.106 & 0.428 & 1200 \\
2459958.6083 & -94.855 & 68.474 & 0.504 & 1200
\enddata
\tablecomments{$h$ is the cross-correlation coefficient \newline}
\end{deluxetable}
\vspace{-0.4cm}

\begin{figure*}[t]
    \centering
    \makebox[\textwidth][c]{\includegraphics[width=1.\textwidth]{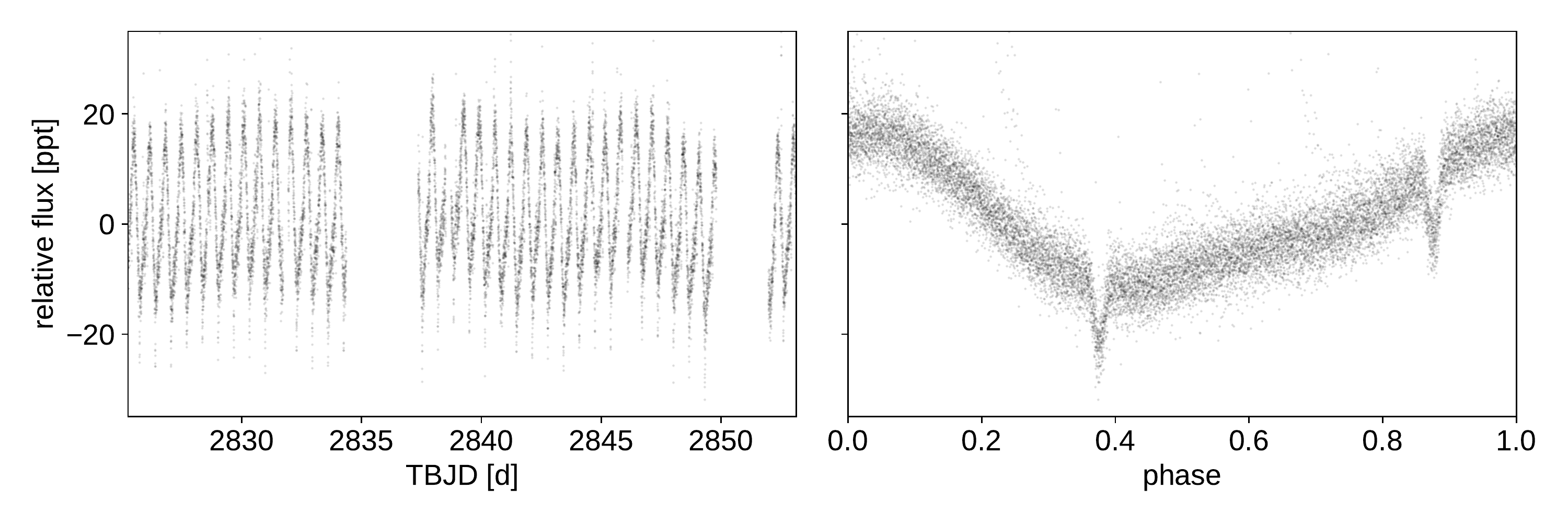}}
    \caption{The TESS sector 56 PDCSAP light curve of G 68-34 AB. The left panel shows the unphased data while the right panel is phased to the 0.655-day rotation period.}
    \label{fig:TESS}
\end{figure*}

\subsection{Photometry}
G 68-34 was observed at 2-minute cadence by TESS \citep{Ricker2015} in its sector 56, which occurred in 2022 September (Julian dates 2459825.25202--2459853.13811). We consider the Pre-search Data Conditioning Simple Aperture Photometry (PDCSAP) light curve provided by the TESS pipeline \citep{Jenkins2016}, from which systematics have been removed using cotrending basis vectors. The light curve exhibits the 0.655-day rotational modulation that we previously identified from MEarth \citep{Nutzman2008, Irwin2015} photometry in \citet{Pass2022}. It also shows grazing primary and secondary eclipses with the same period, indicating that the system is spin--orbit synchronized (Figure~\ref{fig:TESS}). The eclipses were not identified in the MEarth data, as that data set is small (297 observations) and the phased light curve is not precise enough to resolve the small eclipse dips. The eclipses were identified as a threshold-crossing event in the TESS data validation report with a period of 0.32748 days (i.e., half of the binary's orbital period) and a depth of \hbox{11368 $\pm$ 352 ppm}. This report did not flag the system as a suspected eclipsing binary, as the odd/even transit depths are not discrepant at the 2$\sigma$ level. Of course, similar odd/even depths are expected for a roughly equal-mass binary.

To remove flares, we use \texttt{lightkurve} \citep{LightkurveCollaboration2018} to perform a 3$\sigma$ clip of upper outliers from the PDCSAP light curve. We then use the \texttt{flatten} routine to remove the rotational modulation using a Savitzky-Golay filter, identify 3$\sigma$ upper outliers from the flattened light curve, and clip these points from the unflattened PDCSAP light curve as well.

\section{Analysis}
\label{sec:analysis}

We consider two modeling frameworks to derive stellar parameters from our photometric and spectroscopic data. In Section~\ref{sec:model1}, we use \texttt{exoplanet} to perform a preliminary fit modeled after the package's case study on this topic.\footnote{\href{https://gallery.exoplanet.codes/tutorials/eb/}{https://gallery.exoplanet.codes/tutorials/eb/}} This simple model assumes the stars are spherical and neglects information from the phase curve. In Section~\ref{sec:model2}, we perform a more detailed analysis using the \texttt{eb} model presented in \citet{Irwin2011}.\footnote{\href{https://github.com/mdwarfgeek/eb}{https://github.com/mdwarfgeek/eb}} This latter model also allows us to probe the effects of differing spot assumptions on our output stellar parameters. Despite the differing levels of complexity in these models, we ultimately find good agreement between the output stellar parameters.

\subsection{The \texttt{exoplanet} model}
\label{sec:model1}

\begin{figure*}[t]
    \centering
    \makebox[\textwidth][c]{\includegraphics[width=1.\textwidth]{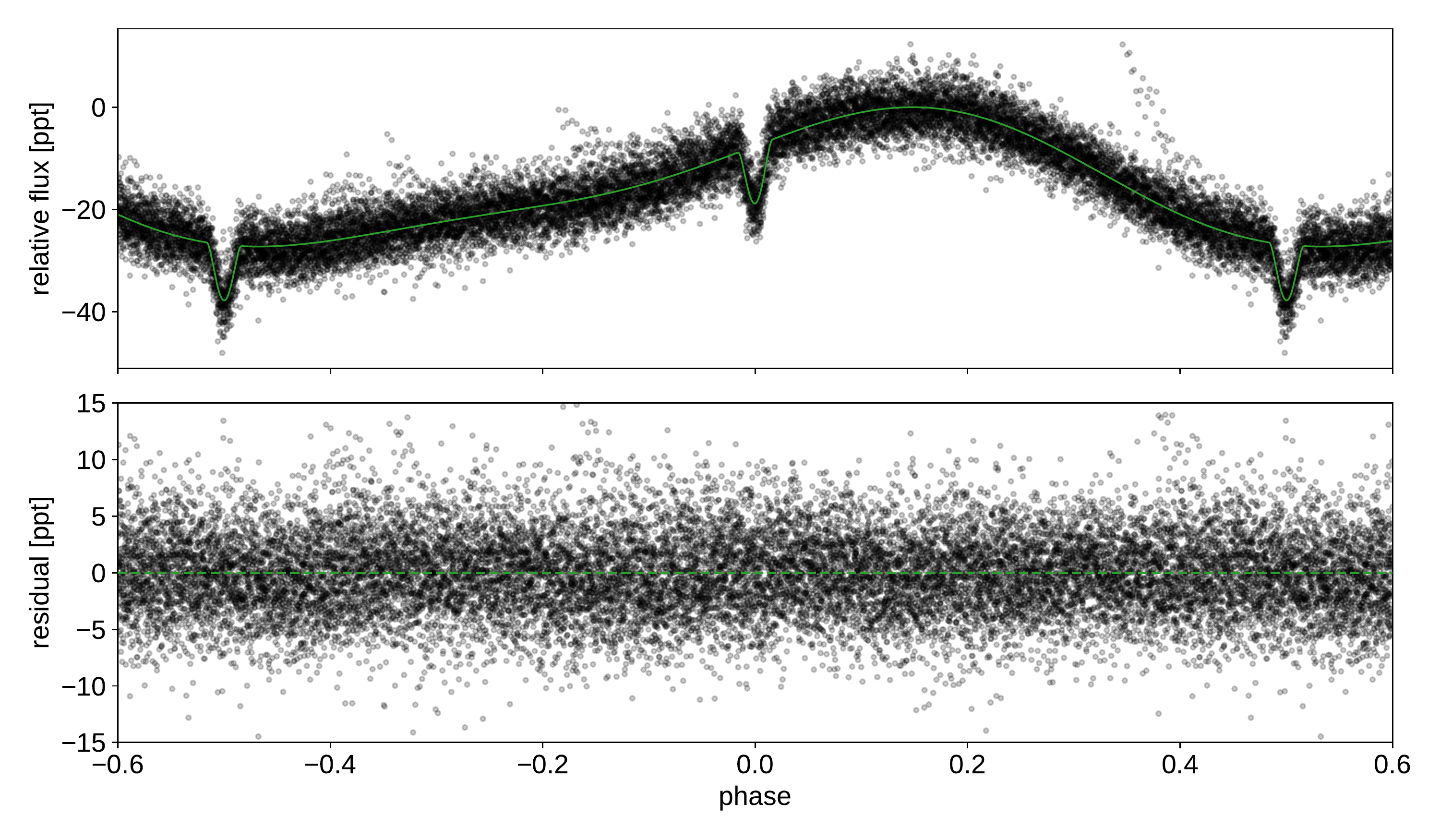}}
    \caption{The photometric model from our \texttt{exoplanet} joint fit.}
    \label{fig:exo-phot}
\end{figure*}

\begin{figure}[t]
    \centering
    \makebox[\columnwidth][c]{\includegraphics[width=1.\columnwidth]{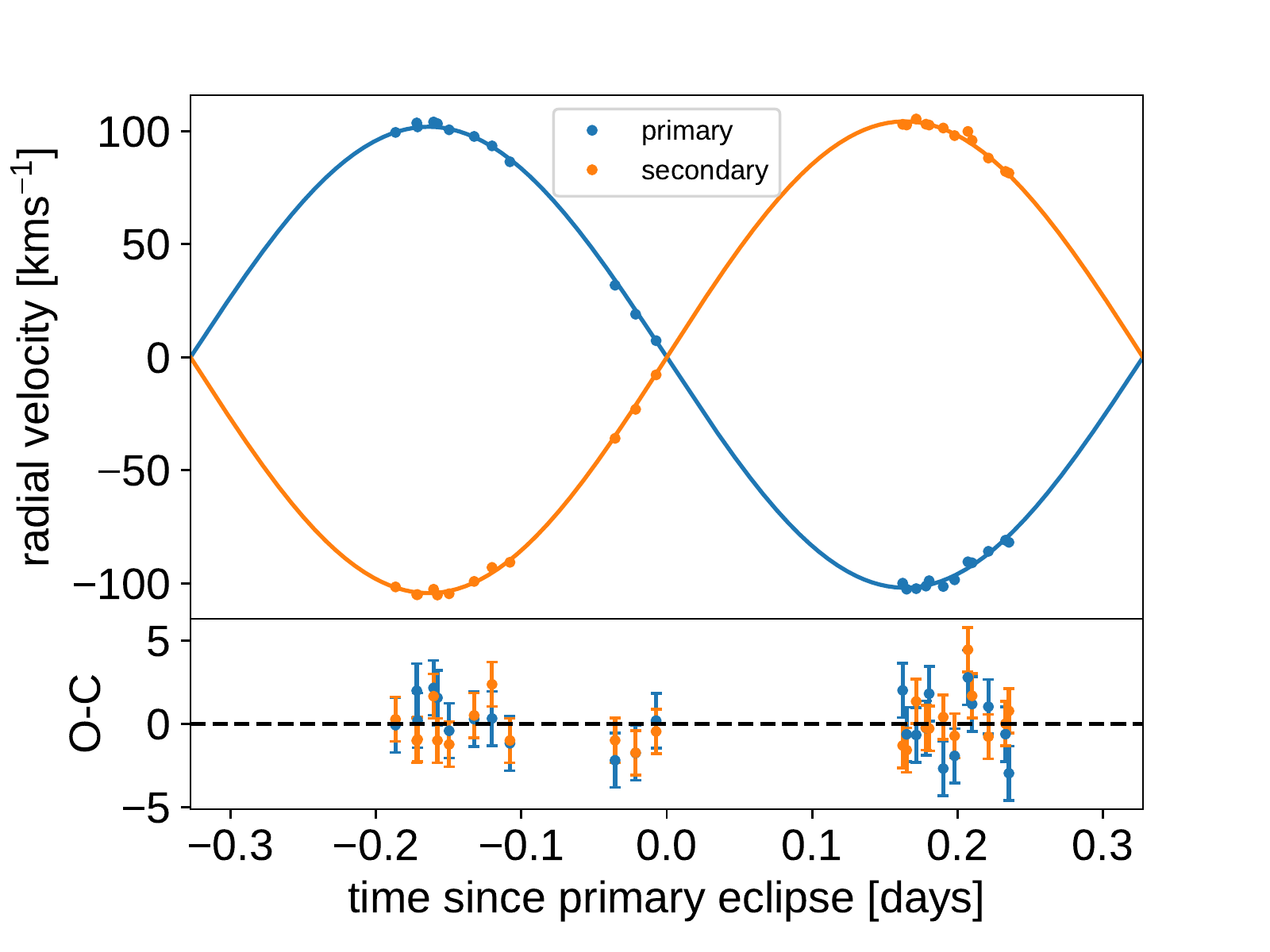}}
    \caption{The radial velocity model from our \texttt{exoplanet} joint fit. From this fit, we estimate typical RV uncertainties of 1.7kms$^{-1}$ for A and 1.4kms$^{-1}$ for B.}
    \label{fig:exo-rv}
\end{figure}

We perform a joint fit with \texttt{exoplanet} using the \texttt{exoplanet.orbits.KeplerianOrbit} framework. Our fit includes a photometric model (Figure~\ref{fig:exo-phot}) that consists of three additive terms: 1) a constant that is a free parameter in the fit; 2) a \texttt{SecondaryEclipseLightCurve} term that models the primary and secondary eclipses using \texttt{starry} \citep{Luger2019}; and 3) a spot model. We also fit a radial velocity model for each star (Figure~\ref{fig:exo-rv}), consisting of two additive terms: 1) a constant that is a free parameter of the fit and that is shared between the two stars; and 2) the radial velocity produced by the \texttt{KeplerianOrbit} model. 

Our spot model consists of sinusoids at the fundamental mode and first harmonic for each star, following the parameterization of Equation 1 in \citet{Hartman2018}. As in Equation 3 of \citet{Irwin2011}, we also include a normalization term such that the spot adjustment is zero at the time of maximum light and spots only serve to reduce the observed flux. As we did not detect a second rotation period in a Lomb-Scargle analysis of the TESS data and we observed similar levels of rotational broadening in the spectra of the two stars, we assume that both stars rotate with a rotation period equal to the orbital period. This model is in good agreement with the observed light curve and we do not find it necessary to include terms for higher-order harmonics. As our spot model is applied as an additive term, we have made the implicit assumption that the spots are on the non-eclipsed portion of the photosphere. We discuss the validity of this assumption in Section~\ref{sec:spot}, when we consider alternate spot architectures.

\begin{deluxetable*}{lccc}[t]
\tabletypesize{\footnotesize}
\tablecolumns{4}
\tablewidth{0pt}
 \tablecaption{Output parameters from our joint RV--photometry fits for G 68-34 A and B.} \label{tab:model_compare}
 \tablehead{
 \colhead{\textbf{Parameter}} & 
 \colhead{\textbf{\texttt{exoplanet} model}} &
 \colhead{\textbf{Irwin+11 \bm{$c_i=0$} model}} &
 \colhead{\textbf{Irwin+11 \bm{$c_i$}-marginalized model}} \\[-0.2cm]
 \colhead{} &
 \colhead{} &
 \colhead{} &
 \colhead{\textit{(preferred)}}}
\startdata
$P$ [days] & 0.654971 $\pm$ 0.000005 & 0.654970 $\pm$ 0.000005 & 0.654970 $\pm$ 0.000005 \\
$a$ [R$_\odot$] & 2.739 $\pm$ 0.008 & 2.742 $\pm$ 0.008 & 2.748 $\pm$ 0.010 \\
$i$ [$\degree$] & 77.15 $\pm$ 0.29 & 77.13 $\pm$ 0.29 & 76.67 $\pm$ 0.52 \\
$T_0$ [day] & 2459825.41585 $\pm$ 0.00014 & 2459825.41586 $\pm$ 0.00014 & 2459825.41586 $\pm$ 0.00014 \\
$R_B/R_A$ & 0.988 $\pm$ 0.040 & 0.980 $\pm$ 0.038 & 0.991 $\pm$ 0.040 \\
$R_A+R_B$ [R$_\odot$] & 0.664 $\pm$ 0.015 & 0.666 $\pm$ 0.015 & 0.686 $\pm$ 0.025 \\
$M_B/M_A$ & 0.977 $\pm$ 0.005 & 0.978 $\pm$ 0.005 & 0.978 $\pm$ 0.005 \\
$K_A + K_B$ [kms$^{-1}$] & 206.3 $\pm$ 0.6 & 206.5 $\pm$ 0.6 & 206.5 $\pm$ 0.6 \\
\textgamma\ [kms$^{-1}$] & -13.01 $\pm$ 0.23 & -13.05 $\pm$ 0.23 & -13.05 $\pm$ 0.23 \\
$M_A$ [M$_\odot$] & 0.3250 $\pm$ 0.0029 & 0.3262 $\pm$ 0.0029 & 0.3280 $\pm$ 0.0034 \\
$M_B$ [M$_\odot$] & 0.3177 $\pm$ 0.0031 & 0.3189 $\pm$ 0.0032 & 0.3207 $\pm$ 0.0036 \\
$R_A$ [R$_\odot$] & 0.334 $\pm$ 0.010 & 0.336 $\pm$ 0.010 & 0.345 $\pm$ 0.014 \\
$R_B$ [R$_\odot$] & 0.330 $\pm$ 0.010 & 0.329 $\pm$ 0.010 & 0.342 $\pm$ 0.014
\enddata
\tablecomments{We recommend using the $c_i$-marginalized estimates, as they account for unknowns in the starspot distribution. \vspace{-0.4cm}}
\end{deluxetable*}
\vspace{-0.9cm}

We assume that the binary orbit is circular, as a circular orbit provides a good fit to the data; if we allow eccentricity to be non-zero, we find that $e<0.006$ with 90\% confidence. Moreover, we would not expect a binary with a 0.655-day orbital period to be eccentric; observations from \citet{Udry2000} indicate that the circularization period for M-dwarf binaries occurs at roughly 10 days.

We use a quadratic limb-darkening law. As the limb-darkening parameters are not well constrained by our grazing eclipses, we require an informative prior. For a $T_{\rm eff}=3300$K, log$g=5.0$ star, \citet{Claret2017} report model limb-darkening coefficients of $u_1=0.1529$ and $u_2=0.4604$ in the TESS bandpass, using the least-square method and the quasi-spherical PHOENIX models. We therefore include a normally distributed prior on the limb-darkening parameters, centered at the \citet{Claret2017} values and with $\sigma=0.05$ to allow for modest differences between the true limb darkening and that predicted by the model.

We also include a normally distributed prior on the flux ratio of 0.94$\pm$0.05 based on the spectroscopic light ratio measured from \texttt{TODCOR}, with the error representing the standard deviation of estimates from each individual spectroscopic epoch. This constraint is important, as it provides most of the information on the radius ratio due to the grazing geometry. The spectroscopic light ratio is measured at similar red-optical wavelengths to the photometry, although the wavelength ranges are not identical. The TRES order covers 7065--7165\AA, while the TESS band extends from 6000--10000\AA\ with an effective wavelength of roughly 8300\AA\ for a model 3300K M dwarf \citep{Allard2012}. As this system is a nearly equal-mass binary, we do not expect that the light ratio would show a strong wavelength dependence. Nevertheless, we also perform our \texttt{TODCOR} analysis on a redder TRES order (order 45, with a central wavelength near 7800\AA) to test the sensitivity of our light ratio to changes in bandpass. We do not find any statistically significant difference in the measured light ratio.

We use uninformative, uniform priors for the other parameters in our model. We also fit for three uncertainty parameters: $\sigma_{\rm RV1}$, the RV uncertainties for A; $\sigma_{\rm RV2}$ the RV uncertainties for B; and $\sigma_{\rm phot}$, the uncertainties in the TESS photometry. We assume each observation has the same error.

We identify the maximum \textit{a posteriori} model using the \texttt{optimize} function in \texttt{exoplanet} and run \texttt{PyMC3} \citep{Salvatier2016} from this starting point, using two chains each with a 1500-draw burn-in and 2000 draws.

\begin{deluxetable*}{llllll}[t]
\tablecolumns{12}
\tablewidth{0pt}
 \tablecaption{Inferred stellar parameters for various longitudinally homogeneous spot distributions \label{tab:spot}}
 \tablehead{
 \colhead{Spot assumption} &
 \colhead{\bm{$i$}} & 
 \colhead{\bm{$M_{\rm A}$}} &
 \colhead{\bm{$M_{\rm B}$}} &
 \colhead{\bm{$R_{\rm A}$}} &  
 \colhead{\bm{$R_{\rm B}$}} \\[-0.2cm]
 \colhead{} &
 \colhead{[\degree]} &
 \colhead{[M$_\odot$]} &
 \colhead{[M$_\odot$]} &
 \colhead{[R$_\odot$]} &
 \colhead{[R$_\odot$]}}
\startdata
$c_{\rm A}=0.0, c_{\rm B}=0.0$ & 77.13 $\pm$ 0.29 & 0.3262 $\pm$ 0.0029 & 0.3189 $\pm$ 0.0032 & 0.336 $\pm$ 0.010 & 0.329 $\pm$ 0.010 \\
$c_{\rm A}=0.0, c_{\rm B}=0.1$ & 76.90 $\pm$ 0.29 & 0.3270 $\pm$ 0.0030 & 0.3197 $\pm$ 0.0032 & 0.332 $\pm$ 0.010 & 0.344 $\pm$ 0.010 \\
$c_{\rm A}=0.1, c_{\rm B}=0.0$ & 76.91 $\pm$ 0.29 & 0.3271 $\pm$ 0.0030 & 0.3198 $\pm$ 0.0032 & 0.349 $\pm$ 0.010 & 0.326 $\pm$ 0.010 \\
$c_{\rm A}=0.1, c_{\rm B}=0.1$ & 76.68 $\pm$ 0.30 & 0.3279 $\pm$ 0.0030 & 0.3206 $\pm$ 0.0032 & 0.344 $\pm$ 0.010 & 0.341 $\pm$ 0.010 \\
$c_{\rm A}=0.0, c_{\rm B}=0.2$ & 76.66 $\pm$ 0.29 & 0.3280 $\pm$ 0.0030 & 0.3208 $\pm$ 0.0033 & 0.328 $\pm$ 0.010 & 0.359 $\pm$ 0.011 \\
$c_{\rm A}=0.2, c_{\rm B}=0.0$ &  76.66 $\pm$ 0.30 & 0.3281 $\pm$ 0.0030 & 0.3208 $\pm$ 0.0032 & 0.365 $\pm$ 0.010 & 0.322 $\pm$ 0.010 \\
$c_{\rm A}=0.2, c_{\rm B}=0.2$ & 76.17 $\pm$ 0.30 & 0.3301 $\pm$ 0.0030 & 0.3228 $\pm$ 0.0033 & 0.357 $\pm$ 0.011 & 0.353 $\pm$ 0.011
\enddata
\end{deluxetable*}
\vspace{-0.8cm}

\subsection{The Irwin et al.\ (2011) model}
\label{sec:model2}
We next consider the \texttt{eb} model described in \citet{Irwin2011}, which is a modified reimplementation of the popular \texttt{JKTEBOP} code \citep{Southworth2004, Southworth2013}. Some small updates to this model are discussed in \citet{Irwin2018}.

While the radial velocity model is similar to that presented in the previous section, the light curve model is more detailed: stars are modeled as spherical during eclipse but as biaxial ellipsoids outside of eclipse, allowing for inclusion of ellipsoidal modulation. As in \citet{Irwin2011}, we fix the gravity darkening parameter at 0.32 \citep{Lucy1967} for both components. As shown in \citet{Torres2021}, this choice is actually inappropriate, as 0.32 is the value of the \textit{exponent} of the gravity-darkening law and not the wavelength-dependent \textit{coefficient}; however, these values are coincidentally quite similar for fully convective  M dwarfs in the TESS band. Moreover, we ultimately find that our fit is insensitive to our choice of this parameter due to the phase curve's domination by spot modulation (Section~\ref{sec:comp}) and therefore we are not overly concerned with its value. We also fix the albedo of each star at 0.4, as in \citet{Irwin2018}, with the reflection effect calculated following \citet{Milne1926} and \citet{Russell1939}.

We adopt the same priors on the light ratio and the limb darkening coefficients as in the previous section. One small difference lies in the treatment of radial velocity uncertainties: while the previous model assumed all observations have the same error, here the points are weighted with respect to the cross-correlation coefficient, as described in Section 4.2 of \citet{Irwin2011}.

For our initial run, we assume that the spots are non-eclipsed and on the primary. However, this framework also contains a flexible spot model that allows us to probe how assumptions about the spot distribution affect our inferred parameters. We discuss these differences in Section~\ref{sec:spot}.

We use the same Monte Carlo parameter estimation as described in \citet{Irwin2011, Irwin2018}: an adaptive Metropolis method with $2\times10^6$ steps, with the first half of the chain discarded as burn-in.

\section{Discussion}
\label{sec:discussion}

\subsection{Comparison of modeling frameworks}
\label{sec:comp}
The first two columns of Table~\ref{tab:model_compare} compare our inferred parameters from the \texttt{exoplanet} and \citet{Irwin2011} fits. The two methods produce results that are fully consistent within uncertainties. This consistency is perhaps unsurprising; while the \texttt{exoplanet} method does not include information from the phase curve, the out-of-eclipse variation is dominated by spot modulation and therefore the phase curve is relatively uninformative. The table also includes a third fit in which we have marginalized over unknowns in the spot distribution using the \citet{Irwin2011} framework. This third set of parameters is our preferred solution and is discussed in more detail below.

\subsection{Influence of spot assumptions}
\label{sec:spot}
We consider each of the spot model variations described in \citet{Irwin2011}. Our fit parameters remain effectively unchanged regardless of whether we explain the observed spot modulation using non-eclipsed spots on the primary, non-eclipsed spots on the secondary, eclipsed spots on the primary, or eclipsed spots on the secondary.

However, one variation can cause a large change in our inferred parameters: the existence of a longitudinally homogeneous distribution of non-eclipsed spots. Such spots do not contribute to the photometric modulation that we observe, as the amount of starlight they block does not change as a function of time. We therefore cannot fit for them; however, if such a component is present, our previous model underestimates the amount of light blocked by the eclipse. This underestimation occurs because we assumed that the region being eclipsed is as bright as the disk-averaged maximum surface brightness. This region is actually brighter, as it is not obstructed by the longitudinally homogeneous spots. In Table~\ref{tab:spot}, we show how our inferred parameters change for various values of $c_i$. Including these spots increases our estimate of the radius of the spotted star and reduces our inferred inclination, slightly increasing our estimate of the component masses. The quantity $c_i$ is described mathematically in \citet{Irwin2011}; in brief, $c_{\rm A}=0.2$ indicates that longitudinally homogeneous spots cause a 20\% decrement in the flux of star A.

We must include an appropriate prior on $c_i$ to marginalize over our ignorance of the true distribution of longitudinally homogeneous spots. But what values of $c_i$ are reasonable for M dwarfs? This quantity is related to the spot filling factor, the fraction of the star's surface area that is covered in spots. However, estimates of M-dwarf filling factors vary dramatically between works. For two active, low-mass M dwarfs studied in \citet{Barnes2015}, the authors found filling factors of only a few percent, while \citet{Jackson2013} infer a typical filling factor of 40\% for an ensemble of magnetically active M dwarfs in the young cluster NGC 2516. In the absence of a predictive framework, we adopt the assumption from \citet{Irwin2011} that values of $c_i$ between 0 and 0.2 are equally plausible. While it is possible that the two stars have vastly different longitudinally homogeneous spot distributions, they both have similar masses, ages, rotation, and H\textalpha\ activity; we therefore further assume that the two stars are more likely to have similar values of $c_i$ than vastly different ones. To approximate this prior, we combine the posterior distributions for the $c_{\rm A}=c_{\rm B}=0.0$, $c_{\rm A}=c_{\rm B}=0.1$, and $c_{\rm A}=c_{\rm B}=0.2$ assumptions, obtaining $c_i$-marginalized estimates of the stellar parameters. This wide prior increases the uncertainty in our radius measurements, as can be seen by comparing the errors associated with the $c_i=0$ model and the $c_i$-marginalized model in Table~\ref{tab:model_compare}. Future work to better understand the filling factors of magnetically active, fully convective M dwarfs will therefore lead to tighter constraints on the properties of this and other eclipsing binary systems.

\subsection{Third light}
We do not include a third light term in the previous models, as the TESS PDCSAP light curve has already been corrected for known contaminants. The contamination ratio ($F_{\rm contam}/F_{*}$) is listed in the TESS Input Catalog \citep{Stassun2019} as 0.58\%; i.e., the applied correction for contamination was small. While overcorrection of sky background has been an issue for some TESS stars \citep[e.g.,][]{Burt2020}, G 68-34 is not located in a crowded region and therefore we do not expect this effect to be significant. Nevertheless, we rerun the $c_i=0$ model with a 1\% third light contribution and find that our fit parameters remain effectively unchanged; modest inaccuracies in the contamination correction will therefore not have an impact on our results. Even a 5\% third light contribution only corresponds to a 1\% offset in our estimated radii and a 0.3\% offset in our estimated masses, which are small relative to the 4\% error in radii and 1\% error in masses we report for our $c_i$-marginalized estimates.

While unknown contaminants are possible, we do not see evidence for their existence. The blended source G 68-34 AB has a 2MASS K-band magnitude of 9.178 mag \citep{Cutri2003} and a Gaia parallax of 25.40 mas \citep{Gaia2022}. Deblending this source into two equal-luminosity components and estimating a mass using the \citet{Benedict2016} K-band mass--luminosity relation yields components masses of 0.32M$_\odot$, consistent with the dynamical masses we have measured. We can repeat this calculation assuming that a 5\% third light contribution is present. We approximate our photometry as $R$ band and use the deblending ratio from \citet{Riedel2014} to convert to a $K$-band light ratio. The third light component would correspond to an M dwarf of mass 0.14M$_\odot$, while G 68-34 A \& B would have masses of 0.29M$_\odot$. These A and B masses are too low to be consistent with our dynamical estimates, considering the 0.014M$_\odot$ uncertainty in the \citet{Benedict2016} relation. The evidence therefore does not support an unknown third component to the 2MASS source contributing significant flux. We also do not see a third set of lines in the spectra, nor a long-term trend in the RV residuals.

\begin{figure*}[t]
    \centering
    \makebox[\textwidth][c]{\hspace{-2cm}\includegraphics[width=1.\textwidth]{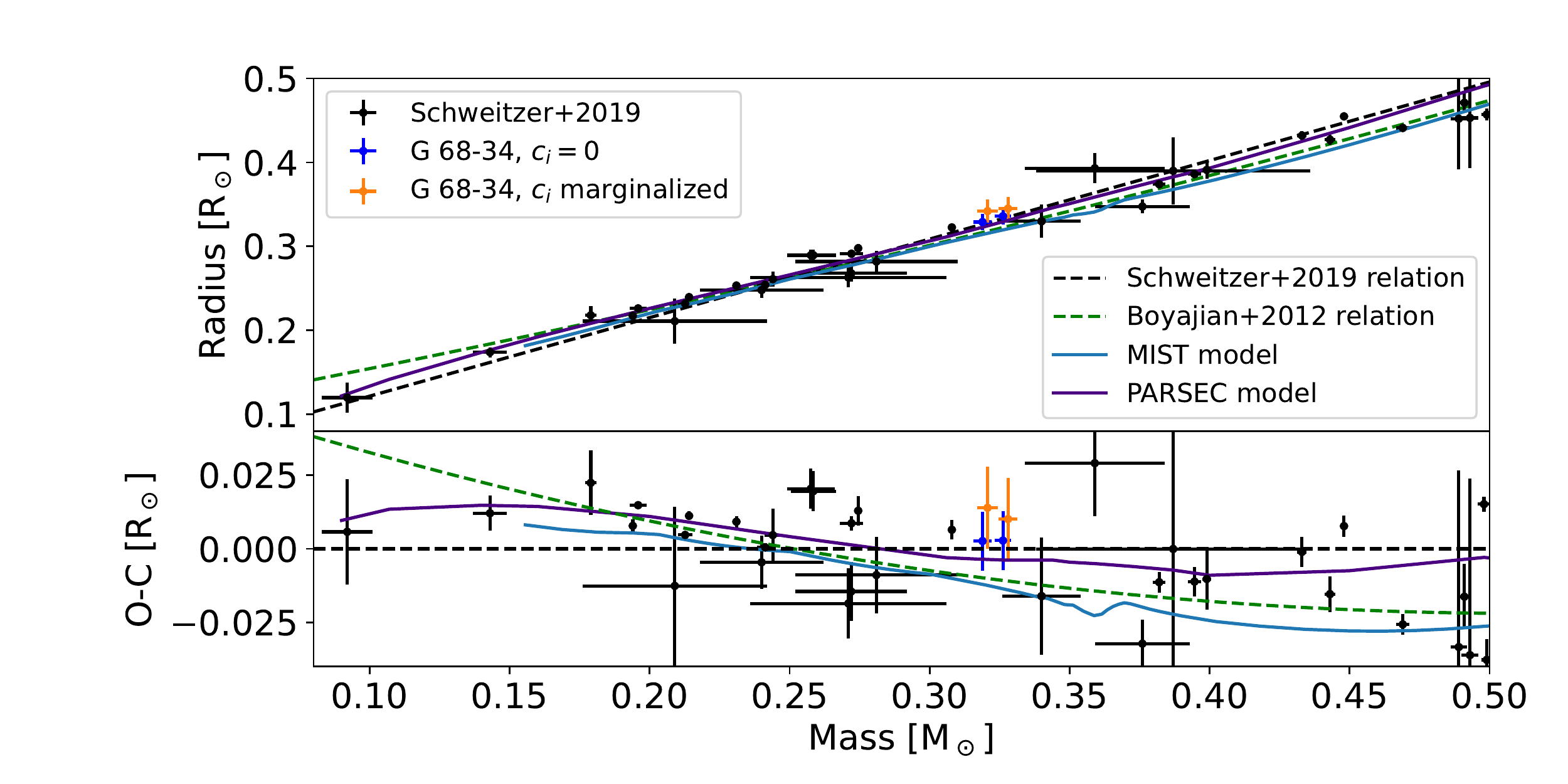}}
    \caption{We compare the masses and radii of the G 68-34 components with M-dwarf DLEBs from the literature, as tabulated in \citet{Schweitzer2019}. The dashed black line is the empirical mass--radius relation from that work. G 68-34 A and B fall along the \citet{Schweitzer2019} relation for $c_i=0$, and slightly above this relation for nonzero values of $c_i$. The green dashed line shows the empirical \citet{Boyajian2012} relation, which uses interferometric radii of single stars. The solid colored lines show two theoretical 5Gyr, [Fe/H]=0 isochrones: from PARSEC in purple \citep{Nguyen2022} and MIST in blue \citep{Choi2016}. In the lower panel, we plot the difference between the observed radii and the empirical \citet{Schweitzer2019} relation.}
    \label{fig:compare}
\end{figure*}

\needspace{6em}
\subsection{Comparison with the literature}
In Figure~\ref{fig:compare}, we compare the masses and radii we determine for G 68-34 A and B to other DLEB M-dwarf systems. G 68-34 A and B are consistent with the empirical mass--radius relation determined by \citet{Schweitzer2019} within uncertainties; however, we note that G 68-34 A and B fall slightly above the relation for nonzero values of $c_i$. The literature M dwarfs with small error bars in the fully convective ($<0.35$M$_\odot$) regime tend to fall similarly above the relation, meaning that the properties of G 68-34 appear normal in the context of other similar systems.

We also show 5 Gyr, solar-metallicity isochrones from two models. We calculate the MIST isochrone \citep{Choi2016} using the MIST web interface\footnote{\href{http://waps.cfa.harvard.edu/MIST/interp_isos.html}{http://waps.cfa.harvard.edu/MIST/interp\_isos.html}, v1.2} and the PARSEC isochrone \citep{Nguyen2022} using the PARSEC CMD input form.\footnote{\href{http://stev.oapd.inaf.it/cgi-bin/cmd}{http://stev.oapd.inaf.it/cgi-bin/cmd}, v3.7} The radius inflation of observed M dwarfs in eclipsing binaries relative to theoretical model predictions has been discussed extensively in past works \citep[e.g.,][]{Torres2013}. Here, we see a significant discrepancy between our observations and the radii predicted by the MIST model, although the PARSEC model is in close agreement with the \citet{Schweitzer2019} relation at the masses of the G 68-34 components, only slightly more than 1$\sigma$ discrepant from our $c_i$-marginalized estimates, and consistent with our $c_i$=0 estimates. While activity has been proposed to explain the discrepancy between the radii of observed M dwarfs in eclipsing binaries and model predictions \citep[e.g.,][]{LopezMorales2007}, works such as \citet{Boyajian2012} and \citet{Mann2015} have argued that the observed inflation of M dwarfs in eclipsing binaries is the result of poor assumptions in the models, as similar inflation is observed in single stars irrespective of metallicity, mass, or stellar activity. \hbox{\citet{Kesseli2018}} and \hbox{\citet{Morrell2019}} similarly concluded that rapidly rotating and slowly rotating M dwarfs experience similar amounts of inflation relative to models. While there is not full consensus on this topic (other works still argue that activity drives radius inflation; e.g., \citealt{Jaehnig2019, Morales2022}), systems like \hbox{G 68-34} with well-characterized properties -- including masses, radii, ages, and activity levels -- are valuable testbeds for models of stellar structure and evolution.

\section{Results and Conclusion}
\label{sec:conclusion}
We discovered that G 68-34 is a double-lined eclipsing binary. Using the high-resolution TRES spectrograph at the FLWO 1.5m telescope, we collected 24 radial velocity observations and performed a joint fit between these radial velocities and photometric monitoring from TESS sector 56. Marginalizing over uncertainties in the starspot distribution, we determined masses to 1\% precision and radii to 4\% precision. Specifically, we found masses of $0.3280\pm 0.0034$M$_\odot$ and $0.3207\pm 0.0036$M$_\odot$ and radii of $0.345\pm 0.014$R$_\odot$ and $0.342\pm 0.014$R$_\odot$. \hbox{G 68-34} is therefore a nearly equal-mass M-dwarf binary, with both components likely being fully convective.

The G 68-34 system is noteworthy in that we can not only measure masses and radii for both components, but also an age due to its widely separated white-dwarf primary, LP 463-28. LP 463-28 has a cooling age of 5.0 Gyr \citep{Pass2022}, meaning that it has existed as a white dwarf for 5.0 Gyr and G 68-34 must be at least this old. To calculate the total age of G 68-34, one must know how long LP 463-28 was on the main sequence before becoming a white dwarf, which is uncertain due to the non-monotonic behavior of the initial-final mass relation for white dwarfs \citep{Marigo2020}. The \hbox{\citet{Cummings2018}} initial-final mass relation suggests a total age of 6.7 Gyr. Following the method described in \citet{Medina2022}, we find that the galactic kinematics of this system is consistent with thin disk membership.

Both G 68-34 A and B exhibit similar levels of rotational broadening. From our TRES spectra, we measured $v$sin$i$ values of 24kms$^{-1}$ for A and 23kms$^{-1}$ for B. The system is spin-orbit synchronized, with the rotational modulation in the photometry sharing the same 0.655-day period as the binary's orbit. If we assume that this synchronization has caused the inclinations of the stars to be the same as the inclination of the orbit, we can estimate $v$sin$i$ based on the radii and inclination inferred in our fit. We find 26.1$\pm$1.1kms$^{-1}$ for A and 25.8$\pm$1.1kms$^{-1}$ for B. These estimates are slightly larger than our nominal spectroscopic $v$sin$i$ measurements. This discrepancy may suggest that low-$c_i$ solutions are favorable, as the predicted $v$sin$i$ is lowered to 25.3$\pm$0.8kms$^{-1}$ and 24.8$\pm$0.8kms$^{-1}$ using the $c_i=0$ solution. However, the errors in our spectroscopic $v$sin$i$ are likely around 2kms$^{-1}$ based on the dispersion we obtain from \texttt{TODCOR} analyses of the individual spectroscopic epochs, and so the data are insufficient to place a compelling constraint on $c_i$.

G 68-34 A and B also exhibit similar levels of H\textalpha\ activity, with equivalent widths of \hbox{-7.1\AA} and \hbox{-6.3\AA}. Systems such as G 68-34 suggest that H\textalpha\ activity is correlated specifically with rotation rate and not with age; old M dwarfs whose rapid rotation persists due to spin--orbit synchronization also retain high levels of H\textalpha\ emission at advanced ages, while single M dwarfs at such ages are found to have H\textalpha\ in absorption.

There are prospects to improve on our mass and radii estimates with future work. Uncertainties could be reduced with a better understanding of some key priors, such as tighter \textit{a priori} constraints on the limb darkening law and the longitudinally homogenous spot distribution. Improved spectroscopic observations could also provide a narrower prior on the light ratio, tightening the constraint on the radius ratio and thereby improving parameter estimates.

\section*{Acknowledgements}
We thank Jonathan Irwin for the development of TRES analysis routines and for providing feedback on this manuscript. We also thank Perry Berlind, Michael Calkins, and Gilbert Esquerdo for collecting TRES observations, Jessica Mink for work on the TRES pipeline, Allyson Bieryla and Sean Moran for operation of the TRES pipeline, David Latham for TRES scheduling, and the anonymous referee for thoughtful suggestions that improved this manuscript. EP is supported in part by a Natural Sciences and Engineering Research Council of Canada (NSERC) Postgraduate Scholarship.

This paper includes data collected by the TESS mission and by the European Space Agency (ESA) mission {\it Gaia} (\url{https://www.cosmos.esa.int/gaia}), processed by the {\it Gaia} Data Processing and Analysis Consortium (DPAC, \url{https://www.cosmos.esa.int/web/gaia/dpac/consortium}). Funding for the TESS mission is provided by the NASA's Science Mission Directorate. Funding for the DPAC has been provided by national institutions, in particular the institutions participating in the {\it Gaia} Multilateral Agreement.

\facilities{FLWO:1.5m (TRES), TESS}
\software{\texttt{eb} \citep{Irwin2011}, \texttt{lightkurve} \citep{LightkurveCollaboration2018}, \texttt{matplotlib} \citep{Hunter2007}, \texttt{numpy} \citep{Harris2020}, \texttt{exoplanet} \citep{ForemanMackey2021, ForemanMackey2021a} and its dependencies \citep{ForemanMackey2017,
ForemanMackey2018, Agol2020, Kumar2019,
AstropyCollaboration2013, AstropyCollaboration2018,
Luger2019, Salvatier2016, TDT2016}.}

\bibliography{sb2}{}
\bibliographystyle{aa_url}



\end{document}